\documentclass[aps,amsmath,pre,preprint,groupedaddress]{revtex4}
\usepackage{graphicx}

\begin{document}


\title{The Phase Behaviors of A Polymer Solution Conf\/ined between Two Concentric Cylinders}


\author{Tongchuan Suo}
\author{Dadong Yan}
\email[Corresponding author:\,]{yandd@iccas.ac.cn}
\affiliation{Beijing National Laboratory for Molecular Sciences
(BNLMS), Institute of Chemistry, Chinese Academy of Sciences,
Beijing, 100190, China}


\date{\today}

\begin{abstract}
A theoretical study on the phase behaviors of a
polymer solution conf\/ined between two coaxial cylindrical walls
is presented. For the case of a neutral inner cylinder, the spinodal point
derived through the Gaussian f\/luctuation theory is conf\/inement-independent
because of the existence of a free dimension in the system.
The kinetic analysis indicates that the fluctuation modes always
have a component of a plane wave along the axial direction, which
can lead to the formation of a periodic-like concentration pattern.
On the other hand, the equilibrium structure of the system is obtained
by using the self-consistent mean-f\/ield theory (SCMFT) and the
interplay between the ``wetting'' phenomenon and the phase separation
is observed by modifying the property of the inner cylinderical wall.
\end{abstract}


\maketitle

\section{introduction}
Rigorously speaking, the classical thermodynamics should be always
used to deal with the systems of inf\/inite particles in an
inif\/inite space. For a real f\/inite system with a large amount of
particles, the classical thermodynamic treatment is valid when the
dimension of the conf\/inement is much larger than the
characteristic length of the system. Otherwise, the boundary of
the system will cause obvious ef\/fects on the interior behaviors.

For a binary f\/luid mixture, an important characteristic length
scale is the correlation length, which will become inf\/inite when
the system approaches the critical state. Hence, the phase
behavior of a conf\/ined binary f\/luid mixture can be affected
much by the boundary of the system. Two common kinds of
conf\/inement are f\/ilm and porous media, which correspond to the
f\/lat boundary case and the curved boundary case, respectively.
The former is always modeled as a system conf\/ined between two
slabs. For this case, when the system is quenched into the
unstable region, the composition f\/luctuations with
characteristic wave vectors parallel to the slabs can grow
unrestrictedly while the ones with perpendicular wave vectors will
be modulated by the conf\/inement. As a result, the microphase
structure in the system must coarsen laterally after the
intermediate stage of the phase separation and the power law of
the domain growth is dif\/ferent from the bulk
case~\cite{jnt.23.1,jcp.113.10386,epl.51.154,polymer.45.2711,polymer.46.977,polymer.46.12004}.
In addition, each slab can be neutral or preferential to a certain
component of the mixture so that the phase behavior can be coupled
with the wetting
phenomenon~\cite{prl.70.2770,jpsb.38.831,prl.96.016107,pre.73.031604,sm.46.12004}.
Compare to the f\/lat boundary case, the porous media is much more
complicated. The simplest model for this case is a binary system
conf\/ined inside a cylindrical pore. Because there is only one
free dimension for the cylindrical pore, the time required by the
system to reach equilibrium is always so long that some metastable
states are long-lived. The wetting behavior also has great
importance to the phase behavior for this
case~\cite{prl.65.1897,pra.44.R7894,pra.46.7664,prl.70.53,pre.50.R4290,pre.52.2736,jcp.122.124510}.

Another interesting case is considered in this work. That is a
polymer solution conf\/ined between two concentric cylindrical
walls. This kind of conf\/inement can be regarded as a combination
of the two that mentioned in the last paragraph and hence is more
general. When the radius of the inner cylinder converges to zero,
the system will reduce to the one in a cylindric pore. On the
other hand, if the radii of the two cylinders converge to
inf\/inite with a constant width between the two walls, the system
will reduce to the one conf\/ined between two slabs. Thus, many
results of this work can be easily generalized to the two limiting
cases. In addition, a polymeric system is always a good object for
studying phase behaviors theoretically because the polymer chain
length is a natural long enough length scale which makes the
coarse-grained treatment and mean-f\/ield theory strictly valid.

In the present work, we give a theoretical study on the phase
behaviors of a polymer solution conf\/ined between two coaxial
cylindrical walls. The self-consistent mean-f\/ield theory (SCMFT)
is used to investigate the equilibrium structure of the system.
The spinodal point of the system is determined by using the
Gaussian f\/luctuation theory. The kinetics of the system are also
studied by combining SCMFT with the Cahn-Hilliard theory.

\section{theoretical framework\label{sec2}}
Consider a polymer solution confined between two inf\/init long
coaxial cylindrical walls, as illustrated in Fig.~\ref{fig1}. The
inner cylinder has the radius $r$ and the outer one has radius
$R$. The average volume fraction of the polymer is
$\overline{\phi}_{\rm P}$, and that of the solvent is
$\overline{\phi}_{\rm S}=1-\overline{\phi}_{\rm P}$. The canonical
ensemble is used here, and the free energy of the system can be
derived by the self-consistent mean-f\/ield
theory~\cite{ma.39.4168,ma.41.5451},
\begin{eqnarray}\label{free energy}
\frac{F}{\rho_{_{\scriptstyle 0}}}&=&\chi\int{\rm d}\mathbf{r}
\phi_{\rm P}(\mathbf{r})\phi_{\rm S}(\mathbf{r})+\int{\rm
d}\mathbf{r} U(\mathbf{r})\phi_{\rm
P}(\mathbf{r})-\sum_{\alpha}\int{\rm d}\mathbf{r}
\omega_{\alpha}(\mathbf{r})\phi_{\alpha}(\mathbf{r})\nonumber\\
&&-\frac{V\overline{\phi}_{\rm P}}{N}\ln\frac{z_{0\rm P}NeQ_{\rm P}}{\rho_{_{\scriptstyle 0}}\overline{\phi}_{\rm P}}
-V\overline{\phi}_{\rm S}\ln\frac{z_{0\rm S}eQ_{\rm S}}{\rho_{_{\scriptstyle 0}}\overline{\phi}_{\rm S}}
\end{eqnarray}
In Eq.~\eqref{free energy}, $\rho_{_{\scriptstyle 0}}$ is the
monomer density, which is def\/ined as monomers per unit volume;
$\alpha=\mbox{P,\,S}$; $\chi$ is the Flory-Huggins parameter,
which quantif\/ies the local interaction between each pair of
polymer segments and solvent molecules;
$\phi_{\alpha}(\mathbf{r})$ is the volume fraction of species
$\alpha$ at point $\mathbf{r}$\,; $\omega_{\alpha}(\mathbf{r})$ is
the auxiliary f\/ield conjuncted to $\phi_{\alpha}(\mathbf{r})$;
$N$ is the degree of polymerization; $z_{0\alpha}$ is the
partition function of component $\alpha$ due to the kinetic
energy, which can be regarded as a constant; $Q_{\alpha}$ is the
partition function of a single molecule of component $\alpha$.
$U(\mathbf{r})$ is the ef\/fective adsorbing potential to the
polymers, which is assumed to be a rectangular-form potential for
simplicity and is just put on the inner cylindrical wall in the
present calculations. It is convenient to use cylindrical
coordinates here and after. Then, $U(\mathbf{r})$ has the form
\begin{equation}\label{ur}
U(\rho,\,z)=\left\{\begin{array}{c@{\quad}l}
-U_0 & r<\rho\leq r+d\\0 & r+d<\rho<R
\end{array}\right.
\end{equation}
where $U_0$ is a constant which quantif\/ies the strength of the
potential; $d$ is the force range of the potential. To obtain the
free energy, we need to solve a set of SCMFT equations,
\begin{gather}
\phi_{\rm P}(\rho,\,z)=\frac{\overline{\phi}_{\rm P}}{NQ_{\rm P}}\int_0^N{\rm d}t
q(\rho,\,z,t)q(\rho,\,z,N-t)\label{pbP}\\
\phi_{\rm S}(\rho,\,z)=\frac{\overline{\phi}_{\rm S}}{Q_{\rm S}}e^{-\omega_{_{\rm S}}(\rho,\,z)}\\
\omega_{_{\scriptstyle \rm P}}(\rho,\,z)=\chi\phi_{\rm S}(\rho,\,z,t)+U(\rho,\,z)+\eta(\rho,\,z)\\
\omega_{_{\scriptstyle \rm S}}(\rho,\,z)=\chi\phi_{\rm P}(\rho,\,z,t)+\eta(\rho,\,z)\\
\phi_{\rm P}(\rho,\,z,t)+\phi_{\rm S}(\rho,\,z,t)=1
\end{gather}
Because the system has a rotational symmetry about the $z$-axis,
the position-dependent functions in the SCMFT equations are
independent on the polar angle. In Eq.~\eqref{pbP}, $t$ is the
coordinate along a polymer chain; the end-integrated propagators,
$q(\rho,\,z,t)$, satisf\/ies the modif\/ied dif\/fusion equation
\begin{equation}\label{mde}
\frac{\partial q(\rho,\,z,t)}{\partial t}=\frac{b^2}{6}\left
[\frac{1}{\rho}\frac{\partial}{\partial\rho} \left
(\rho\frac{\partial}{\partial\rho}\right
)+\frac{\partial^2}{\partial z^2}\right ]
q(\rho,\,z,t)-\omega_{_{\scriptstyle \rm
P}}(\rho,\,z)q(\rho,\,z,t)
\end{equation}
with the initial condition, $q(\rho,\,z,0)=1$. All lengths in the
present calculations are scaled by the Kuhn length $b$ of the
polymer, and thus $b=1$ in Eq.~\eqref{mde}. It should be pointed
out that, to derive the above equations, the well-known Wiener
measure is used to depict the conformation of each polymer chain.
Hence our calculation should be restricted in the weak
conf\/inement case, i.e. $R-r> R_g$ where
$R_g=(\frac{1}{6}N)^{1/2}b$.

According to the Gaussian f\/luctuation
theory~\cite{shi}, when there exists some f\/luctuations around
the mean-f\/ield state, i.e.
$\phi_{\alpha}(\mathbf{r})=\phi_{\alpha}^{(0)}(\mathbf{r})+\delta\phi_{\alpha}(\mathbf{r})$~\footnote{Note
that the f\/luctuation can in principle be dependent on the polar
angle, $\theta$, and so can the two-point correlation functions
considered after.}, the free energy of the system can be written
as an expansion
\begin{equation}
F=F^{(0)}+F^{(1)}+F^{(2)}+\cdots
\end{equation}
where $F^{(0)}$ is the mean-f\/ield free energy illustrated by
Eq.~\eqref{free energy}; $F^{(1)}=0$ since the mean-f\/ield solution
satisf\/ies the SCMFT equations; $F^{(2)}$ has the form as
\begin{equation}\label{F2}
F^{(2)}=\frac{1}{2}\ln\left\{\det\left[\left(\frac{\rho_{_{\scriptstyle
0}}}{\pi}\right)^2C\cdot\tilde{C}\right]\right\}-\ln\int\mathcal{D}\{\delta\phi\}e^{-\mathcal{F}^{(2)}[\{\delta\phi\}]}
\end{equation}
In Eq.~\eqref{F2}, the f\/irst term has no direct relations with
current discussions; the functional
$\mathcal{F}^{(2)}[\{\delta\phi\}]$ in the second term is given by
\begin{equation}\label{mathcalF2}
\mathcal{F}^{(2)}[\{\delta\phi\}]=\frac{\rho_{_{\scriptstyle
0}}}{4}\int{\rm d}\mathbf{r}{\rm d}\mathbf{r}^{\prime}(C^{\rm
RPA})^{-1}(\mathbf{r},\mathbf{r}^{\prime})\delta\phi(\mathbf{r})\delta\phi(\mathbf{r}^{\prime})
\end{equation}
where $\delta\phi(\mathbf{r})=\delta\phi_{\rm
P}(\mathbf{r})-\delta\phi_{\rm S}(\mathbf{r})$. The inverse of the
RPA two-point correlation function in Eq.~\eqref{mathcalF2},
$(C^{\rm RPA})^{-1}(\mathbf{r},\mathbf{r}^{\prime})$, is defined as
\begin{equation}
(C^{\rm RPA})^{-1}(\mathbf{r},\mathbf{r}^{\prime})=\widetilde{C}^{-1}(\mathbf{r},\mathbf{r}^{\prime})
-\chi\delta(\mathbf{r}-\mathbf{r}^{\prime})
\end{equation}
where
\begin{equation}
\widetilde{C}(\mathbf{r},\mathbf{r}^{\prime})=C(\mathbf{r},\mathbf{r}^{\prime})
-\int{\rm d}\mathbf{r}_1{\rm d}\mathbf{r}_2\Delta(\mathbf{r},\mathbf{r}_1)
C^{-1}(\mathbf{r}_1,\mathbf{r}_2)\Delta(\mathbf{r}_2,\mathbf{r}^{\prime})
\end{equation}
In these expressions, the inverse operaters are def\/ined through
the relation $\int{\rm
d}\mathbf{r}_1O^{-1}(\mathbf{r},\mathbf{r}_1)O(\mathbf{r}_1,\mathbf{r}^{\prime})=\delta(\mathbf{r}-\mathbf{r}^{\prime})$;
the formulas of $C(\mathbf{r},\mathbf{r}^{\prime})$ and
$\Delta(\mathbf{r},\mathbf{r}^{\prime})$ can be found in
ref.~\cite{shi}.

Many thermodynamic information of the system can be obtained from
$(C^{\rm RPA})^{-1}(\mathbf{r},\mathbf{r}^{\prime})$.
Particularly, the condition that the smallest eigenvalue of
$(C^{\rm RPA})^{-1}(\mathbf{r},\mathbf{r}^{\prime})$ determines
the spinodal point. In principle, the eigenvalues of $(C^{\rm
RPA})^{-1}(\mathbf{r},\mathbf{r}^{\prime})$ can be derived
by solving the eigenvalue problem
\begin{equation}\label{eigen}
\int{\rm d}\mathbf{r}^{\prime}(C^{\rm
RPA})^{-1}(\mathbf{r},\mathbf{r}^{\prime})\Psi_n(\mathbf{r}^{\prime})=\lambda_n\Psi_n(\mathbf{r})
\end{equation}
where $n$ denotes a complete set of quantum numbers.

\section{results and discussions\label{sec3}}
\subsection{The case of neutral cylinderic walls}
In this section, we mainly focus on the case of neutral
cylindrical walls, i.e. $U_0=0$. Because we always have $R-r\gg b$
in this work, the boundary layers between the bulk of the polymer
solution and the two cylindrical walls are less important. Thus it
is reasonable to omit the boundary layers by letting
$q(\rho,\,z,t)$ satisfy the free boundary condition, i.e. $\left
.\frac{\partial q_{_{\rm P}}}{\partial\rho}\right |_{\rm
boundry}=0$. After these treatments, it can be seen that the SCMFT
equation set has a solution for a homogeneous phase, in which
$\phi_{\rm P}(\rho,\,z)=\overline{\phi}_{\rm P}$ and $\phi_{\rm
S}(\rho,\,z)=\overline{\phi}_{\rm S}$. This trivial solution
represents the equilibrium structure of the system when $\chi$ is
small. However, f\/luctuations will ultimately destroy the
homogeneity of the system with the increasing of $\chi$. As
mentioned above, the critical point at which the system becomes
unstable (i.e. the spinodal point) can be derived from $(C^{\rm
RPA})^{-1}(\mathbf{r},\mathbf{r}^{\prime})$. Rather than solve the
eigenvalue problem in Eq.~\eqref{eigen} directly, it is more
convenient to expand $(C^{\rm
RPA})^{-1}(\mathbf{r},\mathbf{r}^{\prime})$ with the
eigenfunctions of the Laplacian operator, $\nabla^2$. For the
geometry of the system considered here, the eigenfunctions of
$\nabla^2$ are given by
\begin{equation}\label{eigenfunction}
\psi_{nmk}(\rho,\theta,z)=A_{nmk}R_{nm}(\rho)e^{im\theta}e^{ikz}
\end{equation}
with the eigenvalues
\begin{equation}\label{eigenvalue}
\lambda_{nmk}=(\epsilon_n^{|m|})^2+k^2
\end{equation}
where $A_{nmk}$ is the normalization factor; $|m|=0,1,2,\ldots$;
$n=1,2,\ldots$; $k$ is the one-dimensional continuous wave vector
along $z$-direction. In Eq.~\eqref{eigenvalue}, $\epsilon_n^{|m|}$
can be zero or non-zero; when it is non-zero, according to the
boundary condition, it is the $n$-th root of the following
equation
\begin{equation}
\left |\begin{array}{c@{\quad}c}
J_{|m|}^{\prime}(\epsilon_n^{|m|}r) &
N_{|m|}^{\prime}(\epsilon_n^{|m|}r)\\
J_{|m|}^{\prime}(\epsilon_n^{|m|}R) &
N_{|m|}^{\prime}(\epsilon_n^{|m|}R)\end{array}\right |=0
\end{equation}
where $J_{|m|}(x)$ and $N_{|m|}(x)$ are the $|m|$-th order Bessel
function and the $|m|$-th order Neumann function, respectively.
The radial part of the eigenfunction in Eq.~\eqref{eigenfunction},
$R_{nm}(\rho)$, is
\begin{equation}
R_{nm}(\rho)=\left \{\begin{array}{l@{,\quad}l}\delta_{m,0} & \epsilon_n^{|m|}=0\\
J_{|m|}(\epsilon_n^{|m|}\rho)-\frac{J_{|m|}^{\prime}(\epsilon_n^{|m|}r)}{N_{|m|}^{\prime}(\epsilon_n^{|m|}r)}N_{|m|}(\epsilon_n^{|m|}\rho) & \epsilon_n^{|m|}\ne 0\end{array}\right .
\end{equation}
Expanding $(C^{\rm RPA})^{-1}(\mathbf{r},\mathbf{r}^{\prime})$
by $\{\psi_{nmk}\}$ leads to
\begin{equation}\label{CRPA}
(C^{\rm RPA})^{-1}(\mathbf{r},\mathbf{r}^{\prime})=\sum_{nmk}a_{nmk}\psi_{nmk}^*(\mathbf{r}^{\prime})\psi_{nmk}(\mathbf{r})
\end{equation}
The expansion coef\/f\/icient is given by
\begin{equation}\label{anmk}
a_{nmk}=\frac{1}{2\overline{\phi}_{\rm S}}+\frac{\lambda_{nmk}}{4\overline{\phi}_{\rm P}[1-g(\lambda_{nmk}N)]}-\chi
\end{equation}
where $g(x)=\frac{1-e^{-x}}{x}$. As can be seen obviously from
Eq.~\eqref{CRPA}, the eigenfunctions of $(C^{\rm
RPA})^{-1}(\mathbf{r},\mathbf{r}^{\prime})$ are also
$\{\psi_{nmk}\}$ with the eigenvalues ${a_{nmk}}$. Hence, the
spinodal point can be derived from Eq.~\eqref{anmk} that
\begin{eqnarray}
\chi_{_{\rm SP}}&=&\min\left\{\frac{1}{2\overline{\phi}_{\rm S}}+\frac{\lambda_{nmk}}{4\overline{\phi}_{\rm P}[1-g(\lambda_{nmk}N)]}\right\}\nonumber\\
&=&\frac{1}{2}\left(\frac{1}{\overline{\phi}_{\rm S}}+\frac{1}{N\overline{\phi}_{\rm P}}\right)
\end{eqnarray}
Obviously, this result is the same as the non-conf\/ined case and
is consistent with other related works~\cite{jcp.124.144902}.
Physically speaking, the $z$-direction is a free dimension, thus
the system can undergo phase separations along this direction with
no conf\/inement, which is ref\/f\/lected by the
conf\/inement-independent spinodal point.

It is worth noting that the mode corresponding to the minimum
eigenvalue of $(C^{\rm RPA})^{-1}(\mathbf{r},\mathbf{r}^{\prime})$
is a constant, which corresponds a global translation of the
system. This means that if $\chi$ just equals $\chi_{_{\rm SP}}$,
the system is in a critical state but with no phase separation.
However, once $\chi$ exceeds $\chi_{_{\rm SP}}$, many
f\/luctuation modes will be excited to induce the spinodal
decomposition in the system. To study these modes in the
early-stage of the phase separation, we f\/irstly rewrite the free
energy in Eq.~\eqref{free energy} to a functional of $\phi_{\rm
P}(\mathbf{r})$ by using the slow gradient
expansion~\cite{fredrickson2006},
\begin{eqnarray}\label{slow gradient}
F&=&\int{\rm d}\mathbf{r}\left [\chi\phi_{\rm P}(1-\phi_{\rm P})+\frac{\phi_{\rm P}}{N}\ln\phi_{\rm P}
+(1-\phi_{\rm P})\ln(1-\phi_{\rm P})+\frac{1}{36\phi_{\rm P}}(\nabla\phi_{\rm P})^2+\cdots\right ]+{\rm const}\nonumber\\
&\approx&F_0+\int{\rm d}\mathbf{r}\left [(\chi_{_{\rm SP}}-\chi)(\phi_{\rm P}-\overline{\phi}_{\rm P})^2
+\frac{1}{36\overline{\phi}_{\rm P}}(\nabla\phi_{\rm P})^2\right ]
\end{eqnarray}
where $F_0$ is the free energy of the homogeneous state;
the constant $\rho_{_{\scriptstyle 0}}$ is omitted for
simplicity, and the high-order terms are also omitted in the last
step. On the other hand, the time evolution of $\phi_{\rm
P}(\mathbf{r})$ can be depicted by the Cahn-Hilliard
theory~\cite{jcp.28.258,jcp.30.1121,jcp.31.688,jcp.42.93}
\begin{equation}\label{model B}
\frac{\partial\phi_{\rm P}(\mathbf{r},\tau)}{\partial\tau}=\nabla^2\left(\frac{\delta F}{\delta\phi_{\rm P}}\right )
\end{equation}
where $\tau$ is a scaled time variable. By inserting
Eq.~\eqref{slow gradient} into Eq.~\eqref{model B}, it can be
derived that
\begin{equation}\label{C-H eq.}
\frac{\partial\delta\phi_{\rm P}}{\partial\tau}=\nabla^2\left [2(\chi_{_{\rm SP}}-\chi)\delta\phi_{\rm P}
-\frac{1}{18\overline{\phi}_{\rm P}}\nabla^2\delta\phi_{\rm P}\right ]
\end{equation}
where the f\/luctuation $\delta\phi_{\rm P}=\phi_{\rm P}-\overline{\phi}_{\rm P}$
is used instead of $\phi_{\rm P}$.

Then we expand $\delta\phi_{\rm P}$
by $\{\psi_{nmk}\}$. Because of the axial symmetry of the system,
it is only necessary to consider the f\/luctuation modes which
are independent on the polar angle $\theta$. Hence,
\begin{equation}\label{dephi}
\delta\phi_{\rm P}=\sum_{nk}c_{nk}(\tau)A_{n0k}R_{n0}(\rho)e^{ikz}
\end{equation}
By inserting Eq.~\eqref{dephi} into Eq.~\eqref{C-H eq.}, the
evolving equation of the expansion coef\/f\/icient,
$c_{nk}(\tau)$, is derived
\begin{equation}\label{cnk}
\frac{\partial c_{nk}}{\partial\tau}=-2(\chi_{_{\rm SP}}-\chi)\lambda_{n0k}c_{nk}
-\frac{1}{18\overline{\phi}_{\rm P}}\lambda_{n0k}^2c_{nk}
\end{equation}
The solution of Eq.~\eqref{cnk} is
\begin{equation}
c_{nk}(\tau)=c_{nk}(0)e^{\omega_{nk}\tau}
\end{equation}
where
\begin{eqnarray}\label{wnk}
\omega_{nk}&=&2(\chi-\chi_{_{\rm
SP}})\lambda_{n0k}-\frac{1}{18\overline{\phi}_{\rm
P}}\lambda_{n0k}^2\nonumber\\
&=&2(\chi-\chi_{_{\rm
SP}})\left[(\epsilon_n^0)^2+k^2\right]-\frac{1}{18\overline{\phi}_{\rm
P}}\left[(\epsilon_n^0)^2+k^2\right]^2
\end{eqnarray}
When the system is thermodynamically unstable,
$\chi>\chi_{_{SP}}$. Then as can be seen from Eq.~\eqref{wnk}, for
$(\epsilon_n^0)^2+k^2<36\overline{\phi}_{\rm P}(\chi-\chi_{_{\rm
SP}})$, $\omega_{nk}>0$, and the corresponding f\/luctuations
increase with time. In particular, if
\begin{equation}
(\epsilon_n^0)^2+k^2=18\overline{\phi}_{\rm P}(\chi-\chi_{_{\rm SP}})
\end{equation}
$\omega_{nk}$ is maximized, and
\begin{equation}
(\omega_{nk})_{\max}=18\overline{\phi}_{\rm P}(\chi-\chi_{_{\rm SP}})^2
\end{equation}
This value corresponds to a maximum in the rate of increase of
concentration f\/luctuations in the system.

From the discussion above, it can be seen that the possible
f\/luctuation patterns in the unstable system are dependent on
$\overline{\phi}_{\rm P}$, $r$, $R$ and $N$. Generally speaking,
these patterns have two parts, i.e. a radial part and a part along
the $z$-direction, as illustrated by Eq.~\eqref{dephi}. However,
it is interesting that $(\epsilon_1^0)^2$ can be larger than
$36\overline{\phi}_{\rm P}(\chi-\chi_{_{\rm SP}})$ for many
combinations of these parameters. For example, when
$\overline{\phi}_{\rm P}=0.01$, $r=3$, $R=50$ and $N=600$,
$(\epsilon_1^0)^2$ is not smaller than $36\overline{\phi}_{\rm
P}(\chi-\chi_{_{\rm SP}})$ until $\chi$ exceeds 0.8. Note that
$R_0^0(\rho)=1$, then in the above cases, the f\/luctuation modes
that can increase with time must be plane waves along the
$z$-direction. This will lead to a periodic-like concentration
prof\/ile along the $z$-direction in the early stage of the phase
separation.

Although the analysis above is in principle just applicable to the
early stage of the phase separation in the system, it can indeed
be concluded that the phase separation must be along the
$z$-direction even in the late stage. This is because the system
needs to put the interface vertical to $z$-direction in order to
minimize the interface area. As a result, even though the
f\/luctuation mode may have a radial part in the early stage, only
the $z$-direction plane wave part can be maintained as time goes
by.

\subsection{The conf\/ined polymer solution with the presence of an adsorbing potential}
In this section, $U_0$ in Eq.~\eqref{ur} is set to be positive.
Hence, there is an adsorbing potential to the polymers at the
inner cylindric wall. With the presence of this adsorbing
potential, the boundary layer at the inner cylindric wall cannot
be omitted and it needs to use the f\/irst kind boundary condition
for $q(\rho,\,z,t)$, i.e. $q(\rho,\,z,t)|_{\rho=r}=0$, because of
the impenetrability of the inner wall. By contrast, the free
boundary condition can still be used at the outer wall, i.e.
$\left .\frac{\partial q_{_{\rm P}}}{\partial\rho}\right
|_{\rho=R}=0$. Then the equilibrium structure of the system can be
obtained by solving the SCMFT equations numerically. Especially,
Eq.~\eqref{mde} is solved by using the alternating direction
implicit (ADI) method~\cite{jcp.116.7283,pre.74.041808}.

In the present calculations, $\chi$ is always set to be larger
than $\chi_{_{\rm SP}}$. As such, one can expect some interplays
between the adsorption of the polymer by the inner cylindric wall
and the phase separation in the system. The main results are
illustrated in Fig.~\ref{fig2}. Note that the periodic boundary
condition is used along the $z$-direction for convenience in the
calculations, but the period along the $z$-direction in
Fig.~\ref{fig2} is of no physical meanings. The strength of the
adsorbing potential increases gradually from Fig.~\ref{fig2}(a) to
Fig.~\ref{fig2}(c). As can be seen from the f\/igure, when the
adsorbing potential is weak, it has few ef\/fects on the phase
behaviors of the conf\/ined solution, and the system separates
into a polymer rich phase and a solvent rich phase with the
interface being vertical to the $z$-direction. However,
accompanying with the increase of the adsorbing strength, a
``wetting'' layer can be formed by the polymer at the inner
cylindric wall, as illustrated in Fig.~\ref{fig2}(b). Furthermore,
after the strength of the adsorbing potential exceeds a certain
threshold, the interface between the polymer rich phase and the
solvent rich phase becomes along the $z$-direction and then the
concentration profile of the system is independent on $z$
(Fig.~\ref{fig2}(c)).

As been discussed in Sec.~\ref{sec3}A, the interface between the
polymer rich phase and the solvent rich phase prefers to be
vertical to $z$-direction in order to minimize the interface area.
However, the $z$-independent adsorbing potential, $U$, dislikes
the inhomogeneity along the $z$-direction. The competition between
these two factors results in the variations in the concentration
prof\/iles, which are illustrated in Fig.~\ref{fig2}. This
phenomenon is similar to the ``plug-tube'' transition of a binary
liquid mixture in a cylindrical pore. It is worth pointing out
that the threshold of $U_0$, above which the transition happens,
is always not very large, although it is dependent on the
parameters of the system. In our calculation, the value of this
threshold is of order $0.15k_{\rm B}T$ per monomer. Hence if there
exist certain strong interaction, such as the Coulombic
interaction, between the cylindrical wall and the polymer, it is
reasonable to omit the possible inhomogeneity along the
$z$-direction and the calculation can be reduced to one
dimension~\cite{ma.41.5451}.

Before ending this section, we would like to give some discussions
on the periodic-like concentration patterns that are displayed in
parts a and b in Fig.~\ref{fig2}. Although the periods along the
$z$-direction in parts a and b in Fig.~\ref{fig2} are of no
physical meanings, these periodic-like concentration patterns can
indeed be long-lived due to certain kinetic factors, which is
observed by many experiments~\cite{prl.70.53,prl.70.2770} and
simulations~\cite{pra.46.7664,pre.50.R4290,pre.52.2736}. It has
been demonstrated in Sec.~\ref{sec3}A that a periodic-like
concentration prof\/ile along the $z$-direction can form in the
early stage of the phase separation in the present system. In
principle, this initial state will evolve to the equilibrium state
by the process of domain coarsening. However, once the radial size
of each domain reaches the size of the conf\/inement, the
coarsening process can only be achieved through the diffusion and
the coalescence between adjacent domains along the $z$-direction,
which will leads to a very slow kinetics. As a result, the
periodic-like concentration pattern is long-lived, though is not
the equilibrium state. One interesting point is that, if there is
another faster process along with the domain coarsening, such as
crystallization, then the periodic-like concentration pattern can
be maintained. In reality, the inner cylinder in our model can be
certain kind of cylindric adsorber and the outer cylindrical wall
can be the interface between the adsorbed polymer solution and the
outer environment. Hence the results presented here could be a
hint to interpret the mechanism of the formation of the
shish-kebab structure observed in the f\/ield of polymer
crystallization~\cite{cps.237.336,jacs.128.1692}.

\section{summary}
In this paper, we give a theoretical study on the phase behaviors
of a polymer solution conf\/ined between two coaxial cylindrical
walls. The spinodal point of this system, which is derived by using
the Gaussian f\/luctuation theory, is independent on the conf\/inement.
This is because of the existence of the free dimension, i.e. the
$z$-direction. However, the f\/luctuation modes in the system is
greatly af\/fected by the conf\/inement. Due to the kinetic analysis,
the f\/luctuation modes in the early stage of the phase separation
always have a component of a plane wave along the $z$-direction,
which will lead to the formation of a periodic-like concentration
pattern along the axial direction of the system. The equilibrium
structures are obtained by solving the SCMFT equations numerically
and the interplay between the ``wetting'' phenomenon and the phase
separation in the system is also observed. In particular, our results
could give some hints to interpret the mechanism of the formation
of the shish-kebab structure observed in the f\/ield
of polymer crystallization.

\begin{acknowledgments}
T. S. acknowledges Prof. An-Chang Shi for many helpful
discussions. This work is supported by XXXXXXXXX.
\end{acknowledgments}

\bibliography{achemso}

\newpage
\section*{Figure caption}
\begin{figure}
\centerline{\includegraphics[angle=0,scale=1,draft=false]{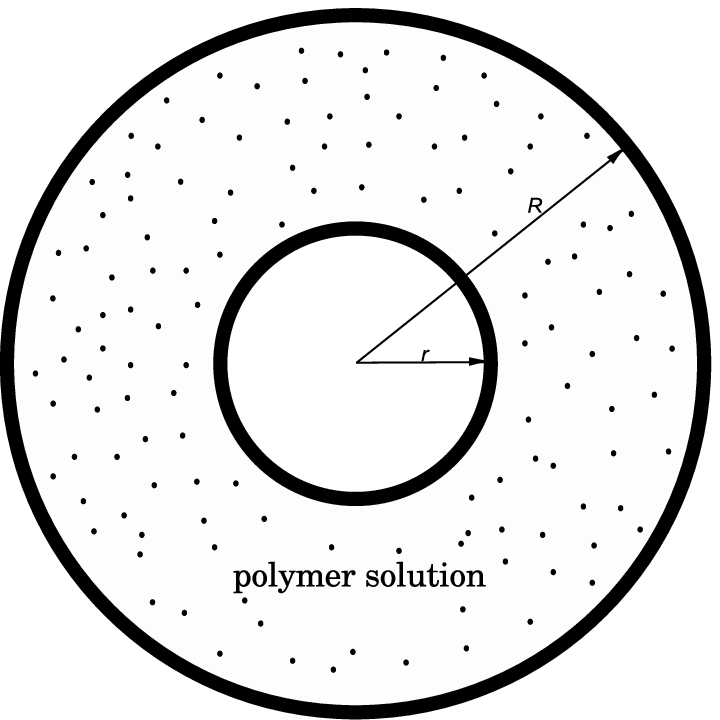}}
\caption{\label{fig1}}
\end{figure}

\textbf{Figure~\ref{fig1}} An illustration of the system
considered.

\bigskip
\textbf{Figure~\ref{fig2}} The concentration profiles obtained by
solving the SCMFT equations with $\chi=0.8$, $N=600$, $r=3$,
$R=100$, $\overline{\phi}_{\rm P}=0.1$ and (a) $U_0=0$; (b)$U_0=0.1$;
(c)$U_0=0.12$.
\newpage
\clearpage


\newpage
\clearpage
\begin{figure}
\centerline{\includegraphics[angle=0,scale=0.6,draft=false]{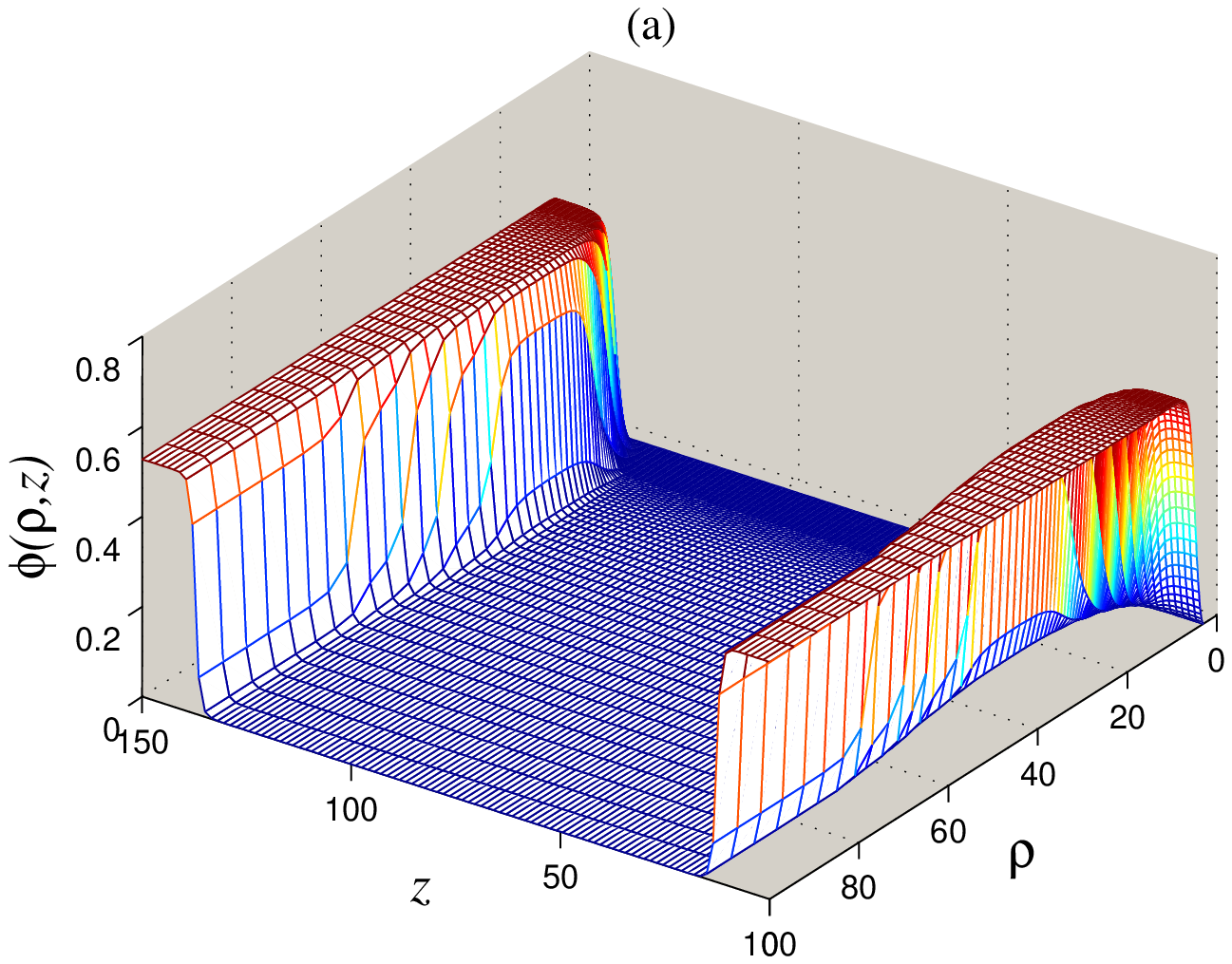}}
\end{figure}
\begin{figure}
\centerline{\includegraphics[angle=0,scale=0.6,draft=false]{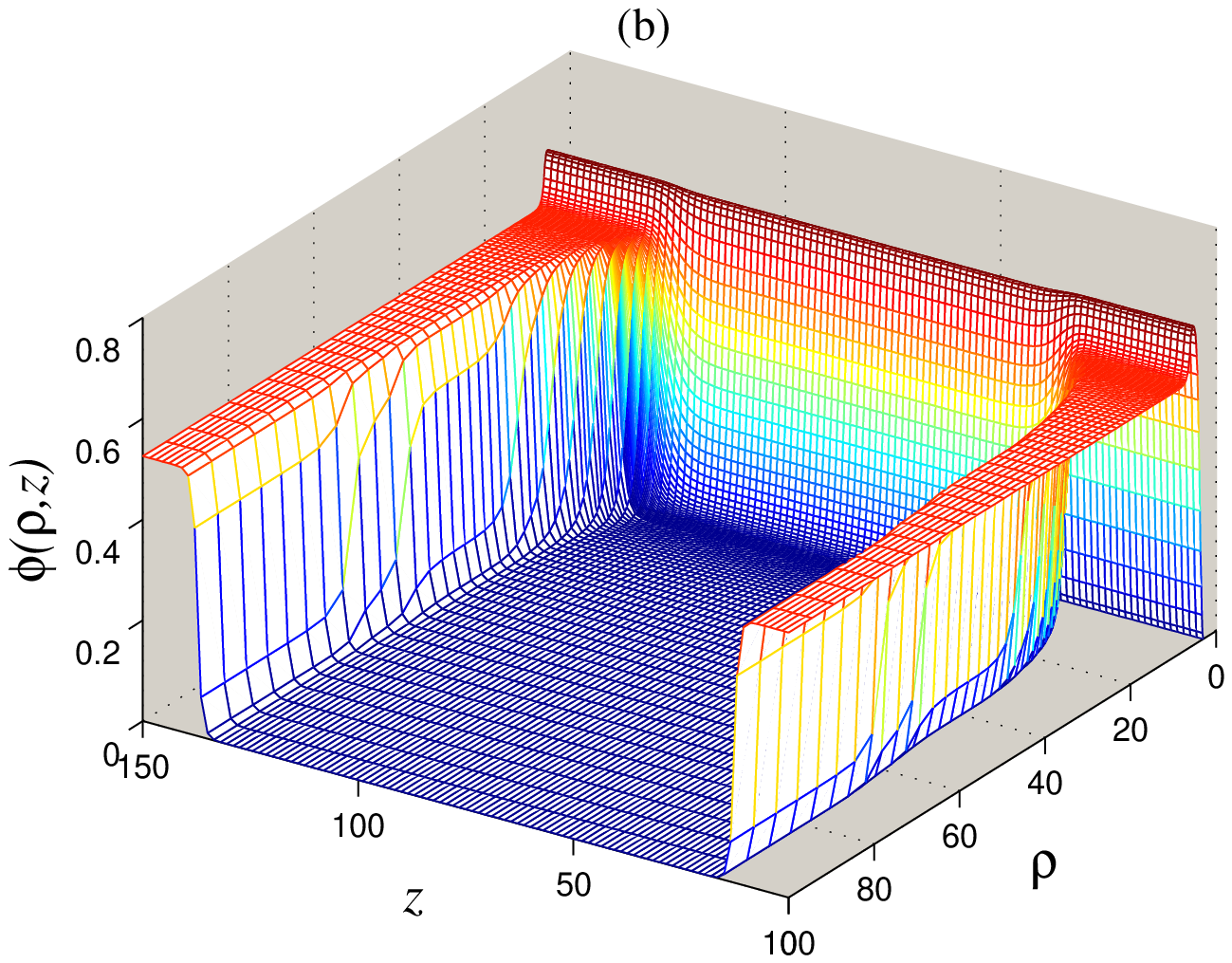}}
\end{figure}
\begin{figure}
\centerline{\includegraphics[angle=0,scale=0.6,draft=false]{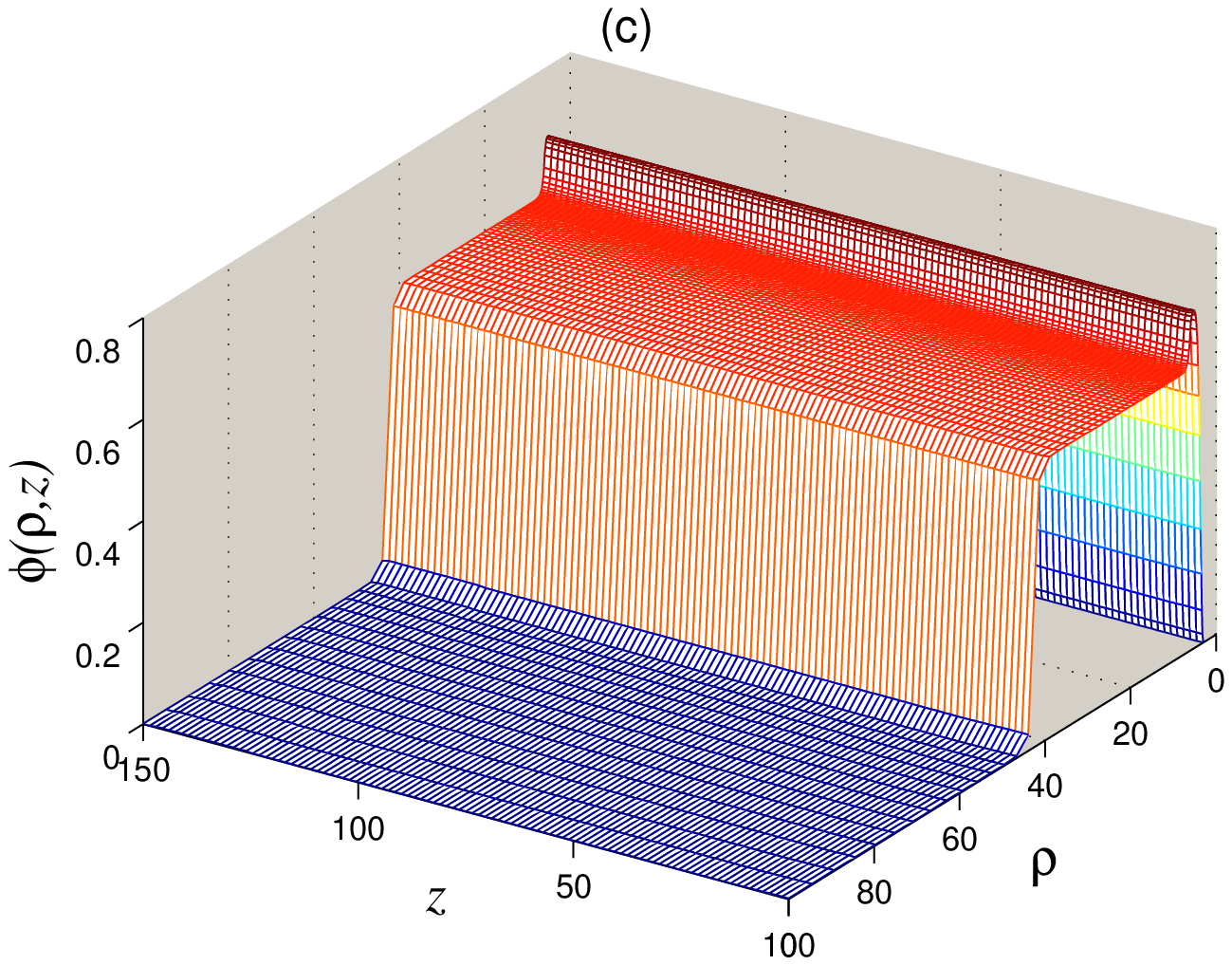}}
\caption{\label{fig2}}
\end{figure}

\end{document}